# Geiger-Mode Avalanche Photodiodes in Particle Detection

E. Vilella[1], O. Alonso[1], J. Trenado[2], A. Vilà[1], M. Vos[3], L. Garrido[2], A. Diéguez[1]

1 – University of Barcelona (UB) – Department of Electronics
C/ Martí i Franquès 1, 08028–Barcelona – Spain

2 – University of Barcelona (UB) – Department of Structure and Constituents of Matter
C/ Martí i Franquès 1, 08028–Barcelona – Spain

3 – Instituto de Física Corpuscular (IFIC)
C/ Catedrático José Beltrán 2, 46980–Paterna – Spain

It is well known that avalanche photodiodes operated in the Geiger mode above the breakdown voltage offer a virtually infinite sensitivity and time accuracy in the picosecond range that can be used for single photon detection. However, their performance in particle detection remains still unexplored. In this contribution, we are going to expose the different steps that we have taken in order to prove the efficiency of Geiger mode avalanche photodiodes in the aforementioned field. In particular, we will present an array of pixels of 1mmx1mm fabricated with a standard CMOS technology for characterization in a test beam.

## 1 Introduction

The International Linear Collider (ILC) [1] and the Compact Linear Collider (CLIC) [2] are two proposed $e^+e^-$ high precision colliders which currently are in their technical design phase. The scope of theses colliders is to provide measurements with unprecedented accuracy and precision. Appropriate detector systems capable of precisely measuring the direction of particle tracks are needed in order to fully exploit the research potential of any particle accelerator. Nevertheless, the future $e^+e^-$ linear colliders put challenging requirements on detector systems since they will have to supply exceptional position resolution at high incoming rates. At present time, there is no mature technology that can fulfill these specifications and new detector systems are being developed in parallel with the accelerator. Solid-state sensor technologies concentrating most of the research, which are based on CMOS monolithic pixel sensors, are Charge Coupled Devices (CCDs) [3], Monolithic Active Pixel Sensors (MAPS) [4], DEPleted Field Effect Transistors (DEPFETs) [5] and Geiger mode Avalanche PhotoDiodes (GAPDs) [6, 7]. Alternative approaches are based on Silicon-On-Insulator (SOI) devices [8]. However, none of the presented candidates meets all the requirements imposed by the collider. Several research groups are currently working towards their improvement. More recently, CMOS sensors exploiting vertical integration technology have also gained interest [9]. This alternative may have the highest potential, but it will need more time to reach maturity.

Although GAPDs' extraordinary capabilities in photon detection are already known [10], their performance in particle detection has not been investigated. To characterize the response of any sensor to high energetic particles, a prototype detector needs to be operated in a test beam environment. In this work, we will present a GAPD detector that has been especially designed to serve as a tracker detector in a test beam. In addition, we will expose the satellite electronics that are also going to be used. The results of the test beam will confirm or refute the validity of the proposed sensor technology as a candidate for tracker detector at the future linear collider.

## 2 Test beam set-up

The set-up for the test beam consists of two GAPD bidimensional arrays (which are the Design Under Test or DUT), two Printed Circuit Boards (PCBs), two FPGAs, an EUDET/AIDA telescope and two scintillator fibres. Additionally, a Trigger Logic Unit (TLU) is used to distribute the trigger signal. A schematic diagram of the test beam set-up is depicted in figure 1.

### 2.1 GAPD detector prototype

The GAPD detector was prototyped with the standard HV-AMS (High-Voltage AustriaMicrosystems) 0.35μm CMOS technology (h35b4). The detector was designed to have a sensitive area of 1mmx1mm in order to facilitate the observation of particle traces. In addition, the sensor size was fixed to 22.9μmx100μm [11] to achieve a good fill factor (88%). The detector is organized in 10 rows (m) per 43 columns (n) of pixels. Each pixel of the detector combines a sensor and proper readout electronics (see figure 2 for pixel schematics). Both the sensor and the readout electronics have been monolithically integrated on a single CMOS die. Given that the scope of the GAPD array is to prove the efficiency of the detector in a test beam, radiation tolerance has not been considered.

The photodiode is implemented by means of a $p^+$/deep n-tub junction, which is surrounded by a p-tub implantation set to prevent premature edge breakdown. Additionally, the corners of the sensor are rounded to avoid electric field peaks at the junction corners. The different pixels within a row share the same deep n-tub (common cathode), which increases the fill factor. Reverse bias overvoltages ($V_{OV}$) over the breakdown voltage ($V_{BD}$) are applied to the sensor cathode to operate the Geiger mode. The readout is performed at the anode or sensing node ($V_S$).

The detector can be operated in the gated mode of acquisition [12] in order to synchronize the sensor operation with the expected signal arrival as well as to reduce the detected noise. In this work, the gated operation is controlled by means of two external signals (RST and INH) implemented through MOS transistors ($M_{N0}$ and $M_{P0}$). When the RST signal is high, and thus the transistor $M_{N0}$ is on, the sensor bias is increased up to $V_{BD}+V_{OV}$ in less than 1ns. As a result, the sensor is recharged and the gated 'on' period is started. On the contrary, when the INH signal is low, and thus the transistor $M_{P0}$ is on, the polarization of the sensor is reduced below $V_{BD}$. The sensor enters the gated 'off' period and it remains in this state until the next rising of the RST signal. When an avalanche is triggered during the gated 'on' periods, the self-sustained current that flows through the junction charges the parasitic capacitance ($C_Q$) of $V_S$ in picoseconds until its voltage raises up to $V_{OV}$. At this point, the polarization of the sensor drops down to $V_{BD}$ and the avalanche is quenched. The node $V_S$ is connected to the readout electronics, which is based on a CMOS inverter ($M_{P1}$ and $M_{N1}$) that converts the analog voltage into a digital pulse. The CMOS inverter was designed to have a threshold voltage of $V_{DD}/2$ and a reduced propagation delay of 50ps. The output of the inverter ($V_{INV}$) is fed in a dynamic latch ($M_{N2}$, $M_{P2}$ and $M_{N3}$), which has been included to make possible in-pixel storage. The dynamic latch is in its sampling mode when the sensor is 'on', whereas the storage intervals are almost coincident with the gated 'off' periods. This performance is controlled by means of the external signal CLK1, which has been implemented through the MOS transistor $M_{N2}$. When the CLK1 signal is set high, which occurs at the same exact time as the RST signal does, the gate $M_{N2}$ is switched on and the dynamic latch enters its sampling mode. The duration of the sampling mode is called period of observation ($t_{obs}$). When the CLK1 signal is set low, the input value of the inverter formed by the transistors $M_{P2}$ and $M_{P3}$ is stored for the gated 'off' period.

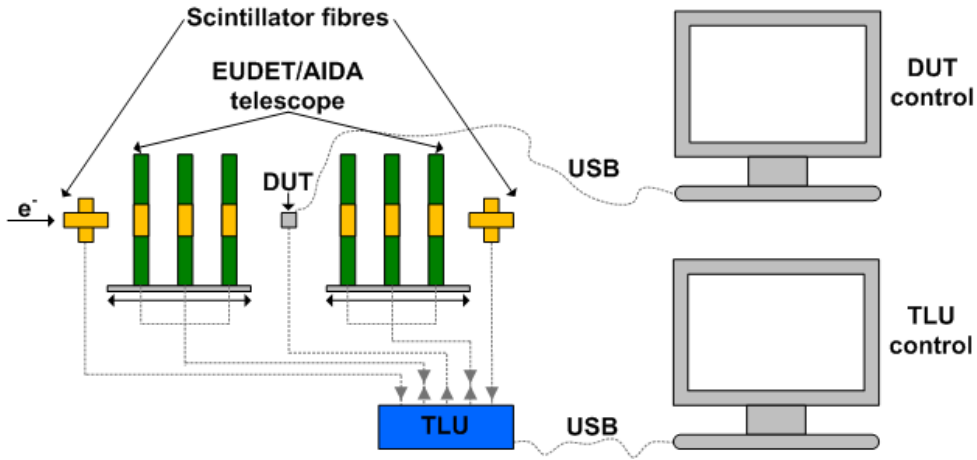

Figure 1: Schematic diagram with the DUT and the satellite electronics for the test beam.

The external signals RST, INH and CLK1 are globally applied to all the pixels simultaneously. The resistance and capacitance of wires in CMOS technologies can introduce delays in the propagation of signals along them. However, the control signals of the detector must reach all the pixels at the same exact time. To estimate the delay in the propagation of the control signals through the distribution wire, a transient simulation with extracted parasitic elements was done. The delay is less than 0.1ns. Hence, no additional elements to compensate the signal propagation delay were included in the circuit.

To increase the Signal-to-Noise Ratio (SNR), a low $V_{OV}$ is desired to obtain a reduced Dark Count Rate (DCR). However, low overvoltages are not allowed in this technology given that the threshold voltage of the nMOS transistors is set at 0.5V. In order to overcome this drawback, the sensor and the readout circuit have different ground nodes, which are respectively GNDA and $V_{SS}$. By raising GNDA with regard to $V_{SS}$, low avalanche voltages can be easily detected by the CMOS inverter.

In order to control the outward data flow, a simple address circuit based on a pass gate ($M_{N4}$) has been placed between the dynamic latch and the output column line. The pass gate $M_{N4}$ is controlled by means of the external signal $CLK2_m$, with m=0 to m=9 (one control signal per row of the detector). The 10 rows of the detector are read sequentially during the gated 'off' periods of the detector. When the $CLK2_m$ signal is set high, the transistor $M_{N4}$ is switched on and the dynamic latch feeds the output column line, which is directly connected to the output pad. Whereas the signal $CLK2_m$ is common to all the pixels of the same row, the output column line n is common to all the pixels of the same column. Consequently, when enabled by the $CLK2_m$ signal, the 43 columns of row m are simultaneously read out. Although this readout configuration needs 43 output pads ($V_{OUTn}$, with n=0 to n=42) plus 13 pads for the control signals (RST, INH, CLK1 and $CLK2_m$, with m=0 to m=9), neither multiplexers nor selection decoders are needed. The whole detector, i.e. the information stored by the 10x43 pixels, can be read in approximately 100ns.

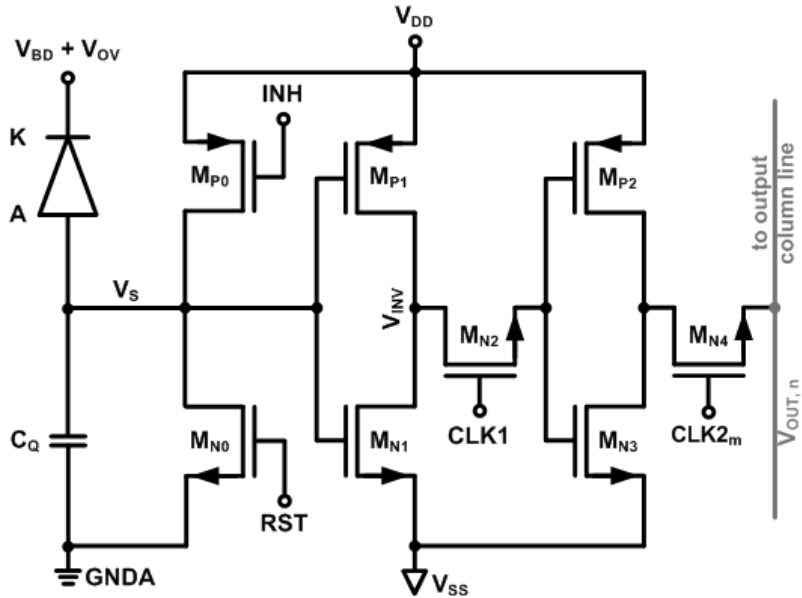

Figure 2: Schematics of the proposed digital pixel.

**2.2 Satellite electronics**

In order to reduce the distortion in the particle path caused by the test set-up materials, the silicon wafer of the chip will be thinned down to 250µm. In addition, the naked die without package will be wire bonded directly to the PCB, which will be perforated under the chip. An ALTERA Cyclone IV FPGA-based control board by terasIC will be used to generate the fast logic control signals (RST, INH, CLK1 and $CLK2_m$) and also to count off-chip the number of pulses generated by the sensor. Two layers of GAPDs will be used to discard the false counts due to the noise.

To characterize the performance of the DUT during the test beam, it is necessary to determine the tracks of the used test beam particles very precisely with a reference system. The resolution of the device used for this purpose should be higher than the expected intrinsic resolution of the DUT. This is usually achieved with beam telescopes, which are placed in the test beam together with the DUT. Thus, it is possible to precisely measure the tracks of incoming particles while studying the response of the DUT. In this work, an EUDET/AIDA telescope [13], which is equipped with MAPS pixel sensors (Mimosa 26), will be used for this purpose. The DUT and the telescope need to be spatially aligned, which will be done with remote-controlled motor stages. Moreover, given that the sensitive area of the telescope is much higher than that of the DUT, another element is needed to discriminate between the hits that occur in the telescope area that corresponds to the DUT from those that occur outside this area. In this work, two scintillator fibres with the same exact area of the DUT will be used for this purpose. One scintillator fibre will be installed in front of the telescope close to the beam source and another one will be placed behind the telescope.

The triggering will be controlled by a TLU [14]. This device generates triggers that are distributed to the EUDET/AIDA telescope and the DUT. It also receives trigger signals from the telescope and the two scintillator fibres. Upon trigger coincidence of the telescope and the two scintillator fibres, the TLU provides the time-stamp. The time-stamp together with the picture of the array (the values of all the pixels, which will be '0' or '1') is either stored in an internal memory or transferred via USB to a PC.

## 3   Expected results

At current time, two GAPD test beams are already planned. The first test beam, which will be a preparation for the final one, will take place at DESY with a 6GeV electron beam. The final test beam will be carried out at CERN with a 120GeV pion beam. In order to know in advance the expected test beam measurement results, simulations have been run with Geant4. To obtain reliable simulation results, the study has been performed including all the different materials that are going to be used in the test beam. Different distances of 2cm and 10cm between the DUT and the last telescope layer have also been considered.

The results of the simulations indicate that there exists a certain amount of distortion associated to the position for which the particle has passed. Since the distortion increases with the distance, it is necessary to place the DUT and the EUDET/AIDA telescope as close as possible. In addition, the expected distortion at DESY is higher than at CERN. In particular, for a DUT-telescope distance of 2cm, the distortion will be about 16µm at DESY and about 0.5µm at CERN. According to these results, and given that the pixel width is 22.9µm, it will be possible to distinguish detection between neighbour pixels at DESY and within a pixel at CERN. Moreover, we also expect to characterize the efficiency of the sensor as a function of the position and the time, the crosstalk, the spatial resolution and the two-track resolution. The measurements will be repeated for different overvoltages.

## 4   Conclusion

A GAPD array has been designed and fabricated with a standard CMOS technology to prove the efficiency of the sensor in a test beam. The detector has 1mmx1mm of sensitive area to fit the test beam requirements. In addition, it can be operated in the gated mode to synchronize the sensor active periods with the expected signal arrival as well as to reduce the detected noise. The set-up will consist of two GAPD arrays, two PCBs, two FPGAs, an EUDET/AIDA telescope, two scintillator fibres and a TLU. At present time, two test beams, at DESY and at CERN, are already planned, where we expect to characterize the GAPD response to high energetic particles. The results of the test beam will confirm or refute the validity of the proposed sensor as a candidate for tracker detector at the future linear collider.

## 5 Acknowledgements

This work has been partially supported by the National Program for Particle Physics through the projects "Desarrollo de nuevas tecnologías en aceleradores y detectores para los futuros colisionadores de Física de Partículas", coded FPA2008-05979-C04-02, and "Desarrollo de nuevos detectores para los Futuros colisionadores en Física de Partículas", coded FPA2010-21549-C04-01. The work has also received funding from the European Commission within Framework Programme 7 Capacities (Grant Agreement 262025) through the project "Advanced European Infrastructures for Detectors and Accelerators – AIDA".